\documentstyle[sprocl]{article}

\input{psfig}

\def\be{\begin{equation}}
\def\ee{\end{equation}}
\def\bea{\begin{eqnarray}}
\def\eea{\end{eqnarray}}

\begin{document}

\title{Semiclassical and quantum production rates}

\author{Piotr Bo\.zek}

\address{ National Superconducting Cyclotron Laboratory,
and Department 
of Physics \\ and Astronomy,
Michigan State University, 
East Lansing, MI-48824\\
and\\ Institute of Nuclear Physics, PL-31-342 Krak\'{o}w, Poland}


\maketitle\abstracts{ We discuss the effects of the off-shell nucleon 
propagation for the particle production. Results from nonequilibrium
Kadanoff-Baym evolution modeling a heavy ion collision are presented.
The production rates are compared to equilibrium emission rates obtained in the
same approximation for the self-energy. The comparison of semiclassical and
quantum in-medium
production rates is performed within the T-matrix approximation.
Self-consistent T-matrix resummation allows us to discuss also the effect of 
the off-shell propagation on the effective in-medium cross section.}

\section{Nonequilibrium evolution}
\label{secn}
In medium modifications of nucleon properties are expected to be important 
at normal nuclear densities and higher. Most of these effects can be taken
 into account by a modification of the effective nucleon mass, cross sections
and mean-field.
 In a strongly interacting medium one expects that off-shell
propagation of nucleons is crucial. On the other hand,
 transport equations using on-shell
quasi-particles proved to be very successful in the description of 
intermediate energy heavy ion collisions. 

Danielewicz in the first work comparing an actual solution of
 the nonequilibrium Kadanoff-Baym
equations with its quasi-particle limit found a slower equilibration in the 
quantum evolution \cite{pawel1}. 
Thus the relaxation rate obtained from the Boltzmann
collision term is larger than obtained from the Kadanoff-Baym collision term, 
including off-shell propagation and memory effects.
The modification of such global transport coefficients due to off-shell
propagation can be taken into account by an effective renormalization of the 
scattering amplitudes between quasi-particles. In a Fermi liquid \cite{pn}
this amounts
 to a multiplication of the cross section by the the factor $z(p_F)^4$
 ($z(p)=(1-\partial {\cal R}e \Sigma(p,\omega)
/\partial\omega|_{\omega=\omega_p})^{-1}$).
 For cold nuclear matter this leads to a reduction 
of the cross section by a factor 2 or more.

Production of very energetic mesons and photons has been observed in heavy
ion reactions. This particle production is subthreshold, i.e. the energy
available in  single nucleon-nucleon collision is insufficient to produce such
an energetic particle. Fermi motion of nucleon inside the colliding nuclei
shifts the effective threshold to higher energies but cannot explain 
the high momentum part of the spectra.
The quasi-particle transport models invoke multiple collisions, or
 pion-nucleon collisions (with pions produced in earlier nucleon-nucleon
 collisions) to explain the production of  energetic mesons or photons.

Subthreshold particle production could be a testing ground for the off-shell 
effects in nuclear propagation. Off-shell spectral function of 
nucleons in hot nuclear matter allows for meson production in the simplest 
one-loop approximation for the  meson self-energy \cite{pb1}.
\begin{figure}[t]
\begin{center}
\psfig{figure=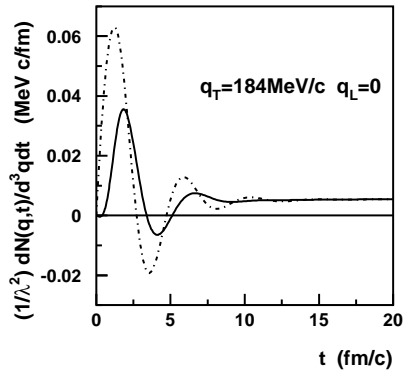,height=5cm}
\end{center}
\caption{Production rate of particles of mass $140$MeV. The solid and 
dashed-dotted lines represent the results with and without imaginary part of 
the contour in the meson self-energy calculation. The Fermion evolution is 
in both cases done with the imaginary part of the contour, i.e with the 
{\it initial correlations}.} \label{fig1}
\end{figure}
Off-shell propagation of nucleons requires the solution of
 evolution equations
for the two-time Green's function, the Kadanoff-Baym equations. 16 years after
the first calculations \cite{pawel1} it is still
limited to spatially homogeneous systems. A numerical solution of a
homogeneous transport equation could be used in the estimate of the particle
production in heavy ion collisions because of the approximate scaling of 
the number of produced particles with the number of nucleons 
participating in the
collision, even for the subthreshold production. The production rate in  
real collision could be estimated by scaling the production rate in 
a homogeneous system by a geometric factor.

In previous work \cite{pb1} we studied the production in
  the nonequilibrium dynamics of 
nucleons. The evolution of the system follows the 
Kadanoff-Baym equations, in the direct second 
Born approximation
\begin{equation}
\left( i\frac{\partial}{\partial t_1} -\omega_p\right)G^{<(>)}(p,t_1,t_2)=
\int_C dt^{'} \Sigma(p,t_1,t^{'}) G(p,t^{'},t_2) \ . \end{equation}
The integration contour $C$ involves an imaginary part giving,
within the second Born approximation, the ground state of nuclear matter
with two boosted Fermi spheres. The real part of the contour gives the 
collision integral with memory. In order to get the spectral function far
from the quasi-particle pole, small steps in times $t_1,t_2$ have to be used
 \cite{pb1}.

The meson production rate is obtained from the one-loop meson 
self-energy \cite{pb1}
\begin{equation}
\frac{dN(p,t)}{d^3pdt}=2{\cal R}e\left(-\int_{t_0}^t+\int_{t_0}^{t_0-i\tau_0}
dt^{'} \Pi^<(p,t,t^{'})D^>_0(p,t^{'},t) \right) \ , 
\end{equation}
where
\begin{equation}
\Pi^<(p,t_1,t_2)=-i\lambda^2\int\frac{d^3q}{(2 \pi)^3} G^<(p-q,t_1,t_2)
G^>(p,t_2,t_1)
\end{equation}
and $D_0$ is the vacuum meson propagator (we neglect the influence of the
meson-fermion interaction on the nucleon evolution and we calculate only the 
production  rate not the full meson transport equation).
The results in Fig.1 show that the production rate is oscillating with large 
amplitude at initial times. It also changes sign. It is a manifestation of the 
oscillating time exponent, which gives the energy conservation in the limit 
$t_0\rightarrow - \infty $. In an interacting system with off-shell nucleons
 the positive 
oscillations dominate giving a finite production rate from one-loop diagram. 
It is however not possible to define a particle production rate at  initial
times. The initial correlations take into account only the fermion-fermion
 interaction via a two-body potential. There is no initial meson content in the
collision, i.e. the initial state is not correlated with respect to the 
meson-fermion coupling. This and the previously discussed meson formation time 
(oscillations of the rate) dominate the meson production at initial times. The
meson production rate can be defined only at large times $t\simeq 20$fm/c
where the nucleon sector is basically equilibrated.

\section{Equilibrium rates}
Production rates in equilibrium finite temperature nuclear matter 
can be obtained with much less numerical effort
and higher accuracy \cite{pb2}.
Equilibrium spectral functions for nucleons, at finite temperature and 
density, are obtained by iterative 
solution of the set of equations
\begin{eqnarray}
\label{iteq} & &
\pm iG^{<[>]}(p,\omega)=A(p,\omega) f(\omega)\left[ (1-f(\omega)) \right]
\ , \nonumber \\ & &
A(p,\omega)=\frac{-2{\cal I}m\Sigma(p,\omega)}{(\omega -p^2/2m -{\cal R}e
\Sigma(p,\omega))^2+{\cal I}m\Sigma(p,\omega)^2} \ ,\nonumber \\ & &
\Sigma^{<[>]}(p,\omega)={\cal F}(G^<,G^>) \ ,\nonumber \\ & &
{\cal R}e \Sigma(p,\omega)=\int\frac{d\omega^{'}}{\pi}\frac{{\cal I}m
\Sigma(p,\omega^{'})}{\omega^{'}-\omega} \ ,\nonumber \\ & &
-i\int \frac{d^3pd\omega}{(2 \pi)^4} G^<(p,\omega)=\rho  \ .\nonumber
\end{eqnarray}
The functional  ${\cal F}$ in the equation for $\Sigma^{<[>]}$ corresponds
to the direct second Born approximation as in the nonequilibrium evolution.
In Fig.2 we present the spectral function obtained at $T=10$MeV. The spectral 
function is broad far from the Fermi momentum, but gets narrower for momenta
 close to it. As expected for a Fermi liquid 
the spectral function at the Fermi momentum
gets very peaked when decreasing the temperature.

The quantum production rate of mesons with momentum 
$p$ and energy $\Omega_p=\sqrt{p^2+M^2}$
in the one-loop approximation for the meson 
self-energy is 
\begin{equation} 
\frac{dN(p)}{d^3pdt}=4\lambda^2\int\frac{d^3q d\omega}{(2 \pi)^4}
A(q,\omega)A(q-p,\omega-\Omega_p)(1-f(\omega))f(\omega-\Omega_p) \ .
\end{equation}
The  semiclassical production rate
is 
\begin{eqnarray}
\frac{dN(p)}{d^3pdt}&=&\int \frac{d^3p_1}{(2 \pi)^3} \dots \frac{d^3p_4}
{(2 \pi)^3} |M|^2 (2 \pi)^3 \delta(\sum p_i -p) 2 \pi \delta(\sum \omega_i-
\Omega_p) \nonumber \\ & & f(p_1)\dots (1-f(p_4)) \ , \end{eqnarray}
with the matrix element for particle emission given by
\begin{eqnarray}
\label{mat}
|M|^2=4 \lambda^2 V(p_2-p_4)^2 & \bigg( &|G^+(\omega_{p_1}-\Omega_q,p_1-q)|^2
\nonumber \\ & +&
|G^+(\omega_{p_3}+\Omega_q,p_3+q)|^2 \bigg) \ .
\end{eqnarray}
This matrix element is obtained by including self-energy corrections in the 
one-loop quasi-particle meson self-energy.
The self-energy in the retarded propagators in $|M|^2$ as well as the single
particle energies $\omega_i$ are taken from the quantum calculation of finite 
temperature nuclear matter.

\begin{figure}[t]
\parbox[t]{5.5cm}{
\psfig{figure=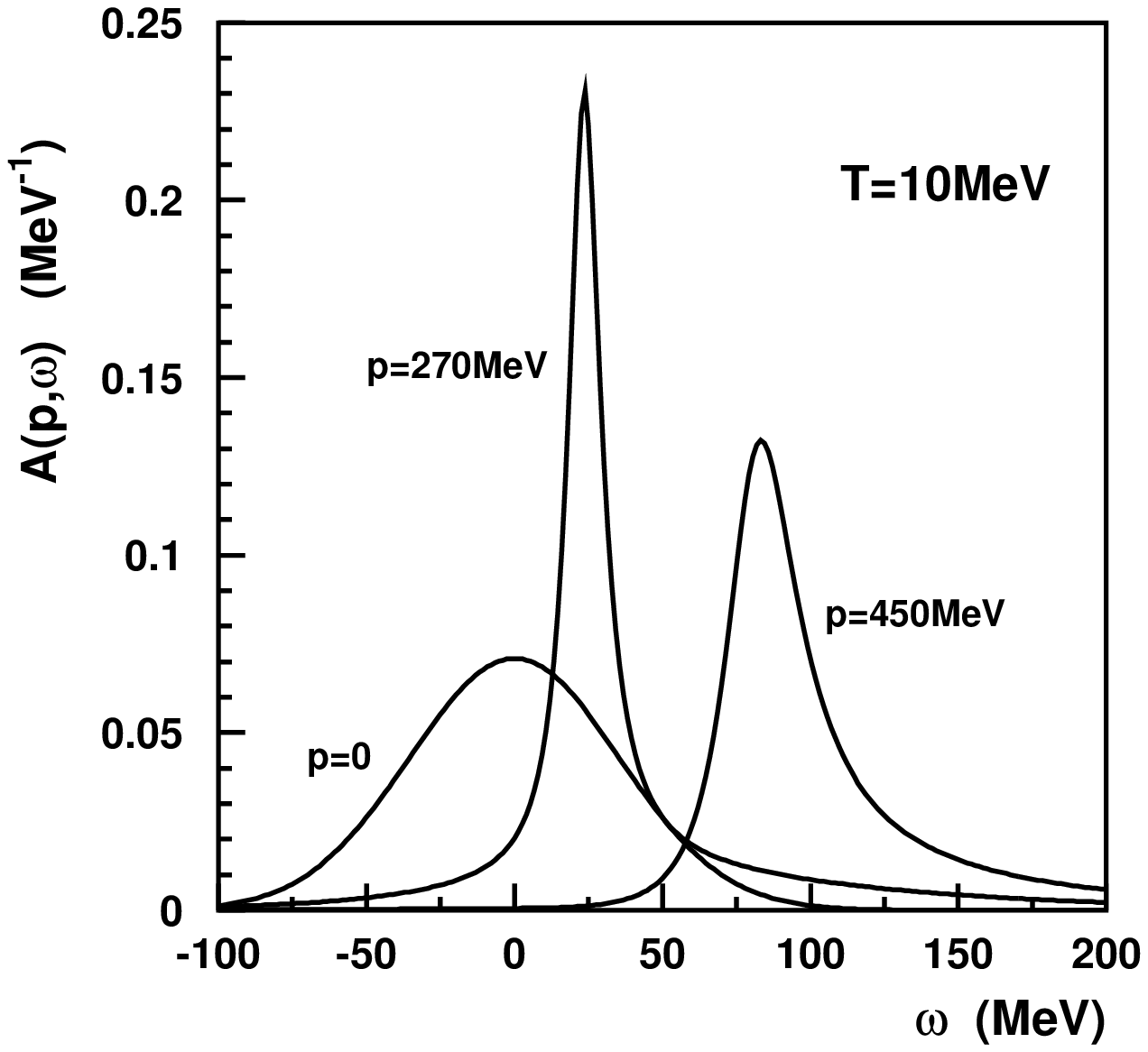,width=5cm,height=5cm,angle=0}}
\hspace{3ex}
\parbox[t]{5.5cm}{
\psfig{figure=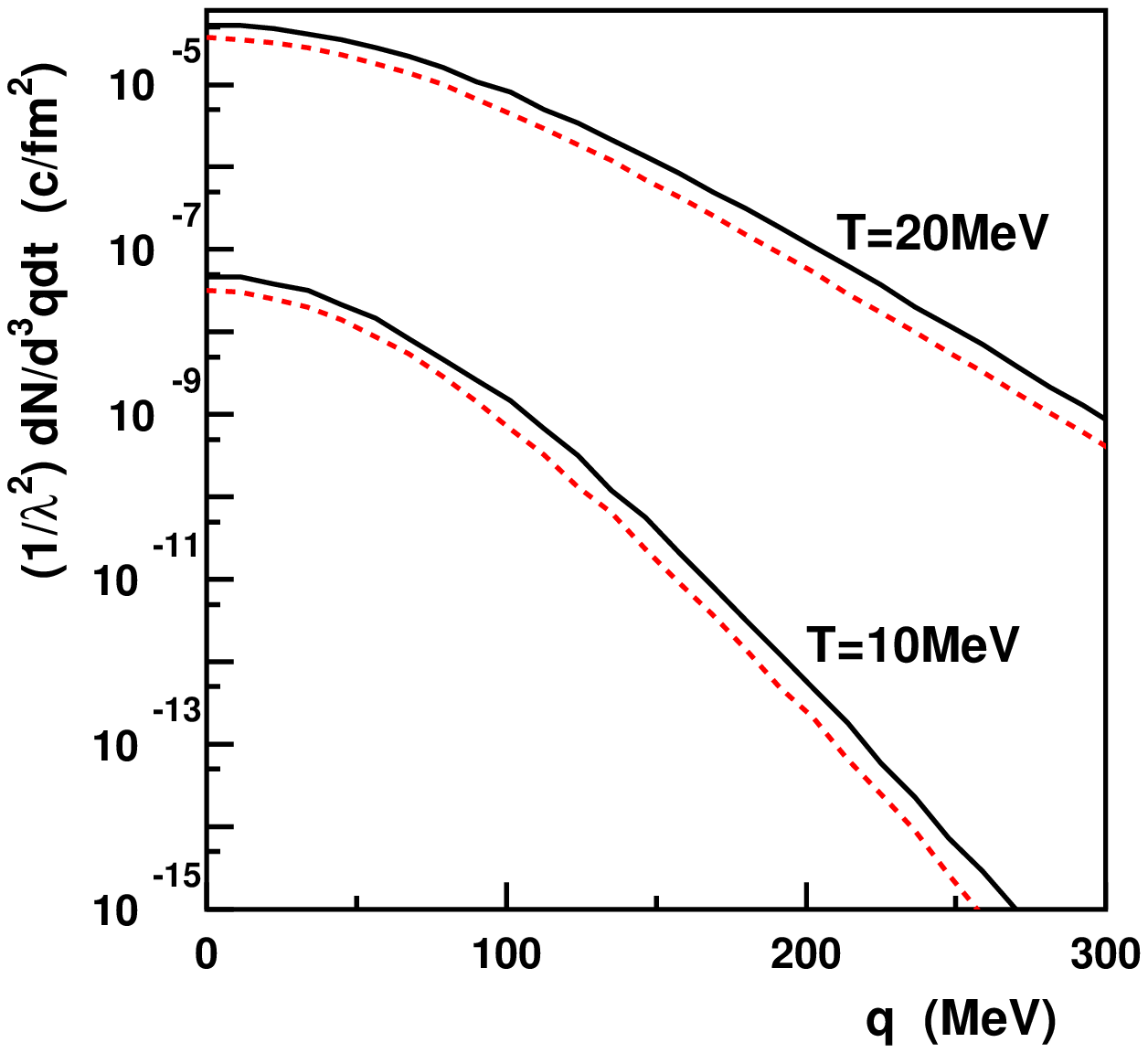,width=5cm,height=5cm,angle=0}}
\caption{ (left panel) Nucleon spectral function at normal nuclear 
density and $T=10$MeV for several momenta.
(right panel) Production rate of mesons of mass 
$140$~MeV at equilibrium.
 Quantum and  semiclassical production rates are represented by solid and 
dashed lines respectively.}\label{figb}
\end{figure}

The production rate for particles of mass $140$MeV is $\sim 2$ times larger
in the quantum calculation. The self-consistent spectral function
involves processes with multiple sequential nucleon-nucleon
 collisions which populate
 far off-shell regions of the spectral function.
Also the one-loop self-energy with off-shell fermion propagators includes 
contribution from 3-nucleon collisions corresponding to self-energy insertions 
on both propagators in the loop. These contribution are not taken into account
 by the 2-body matrix element (Eq. \ref{mat}).
The difference between the semiclassical and quantum production rates 
increases with the energy of the produced particle. On the other hand, 
the emission of soft 
particles is stronger in the semiclassical calculation. All the emission 
rates increase strongly with temperature. 
Quantum production rates are similar as obtained from the nonequilibrium 
evolution \cite{pb1}
 (Sec. \ref{secn}) at large times. The semiclassical rates published
 in the work \cite{pb1} are different and obviously wrong.
The overall collision rate is larger in the semiclassical collision integral,
similarly as in the nonequilibrium calculation.

\section{T-matrix approximation}
In this section we discuss   in-medium modification of the
 cross-sections.  This requires the use of the T-matrix approximation instead 
of the Born self-energy. T-matrix ladder resummation with off-shell
propagators has only recently been
achieved \cite{haus}. The equation for the T-matrix is
\begin{eqnarray}
\label{teq}
<{\bf p}|T^\pm({\bf P},\omega)|{\bf p}^{'}>& =& V({\bf p},{\bf p}^{'})+ 
 \int\frac{d^3k}{(2 \pi)^3}
\int\frac{d^3q}{(2 \pi)^3} V({\bf p},{\bf k}) \nonumber \\ 
& &<{\bf k}|{\cal G}^\pm({\bf P},\omega)|{\bf q}>
 <{\bf q}|T^\pm({\bf P},\omega)
|{\bf p}^{'}> \ ,
\end{eqnarray}
\begin{figure}
\parbox[t]{5.5cm}{
\psfig{figure=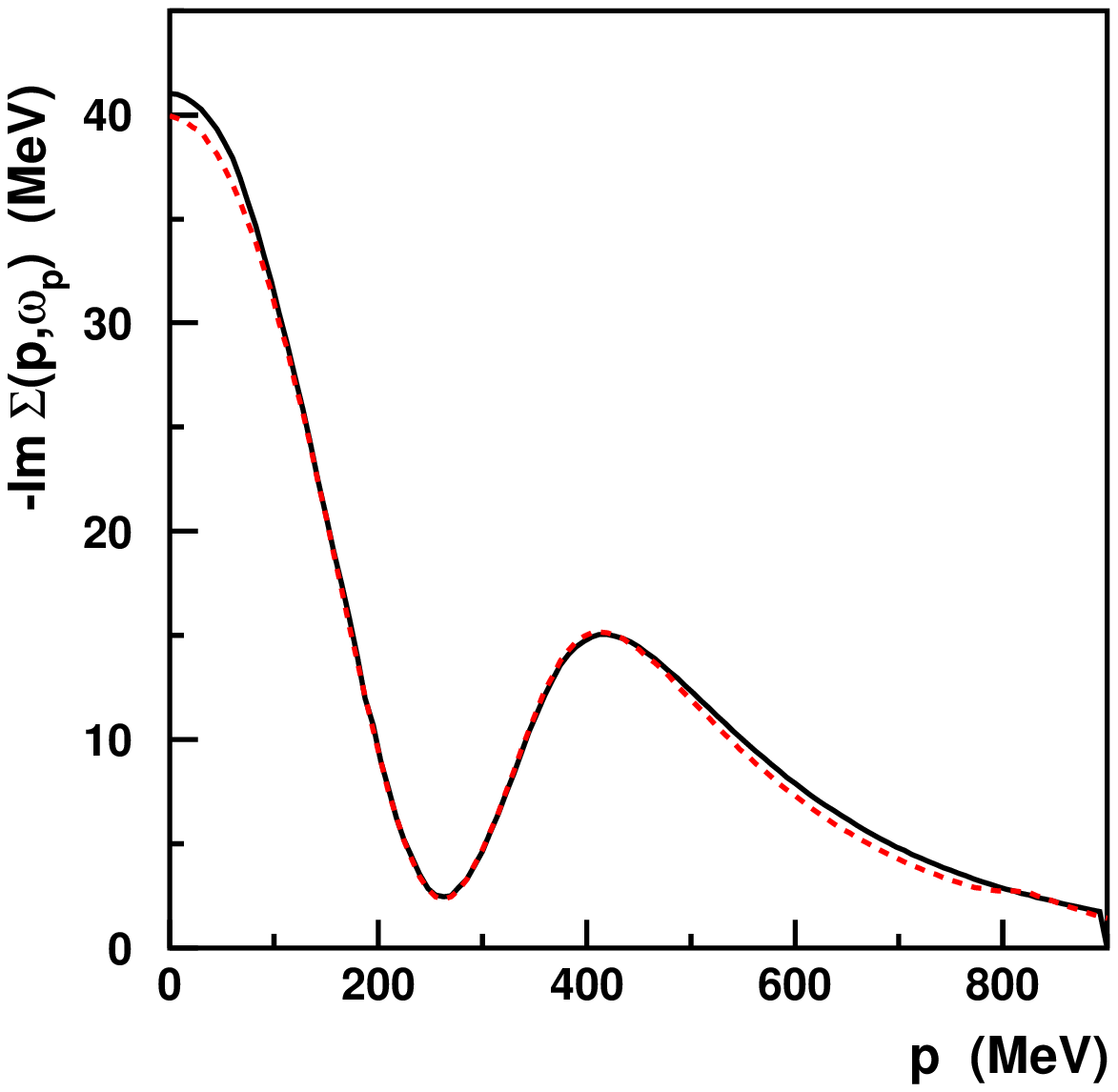,width=5cm,height=5cm,angle=0}}
\hspace{3ex}
\parbox[t]{5.5cm}{
\psfig{figure=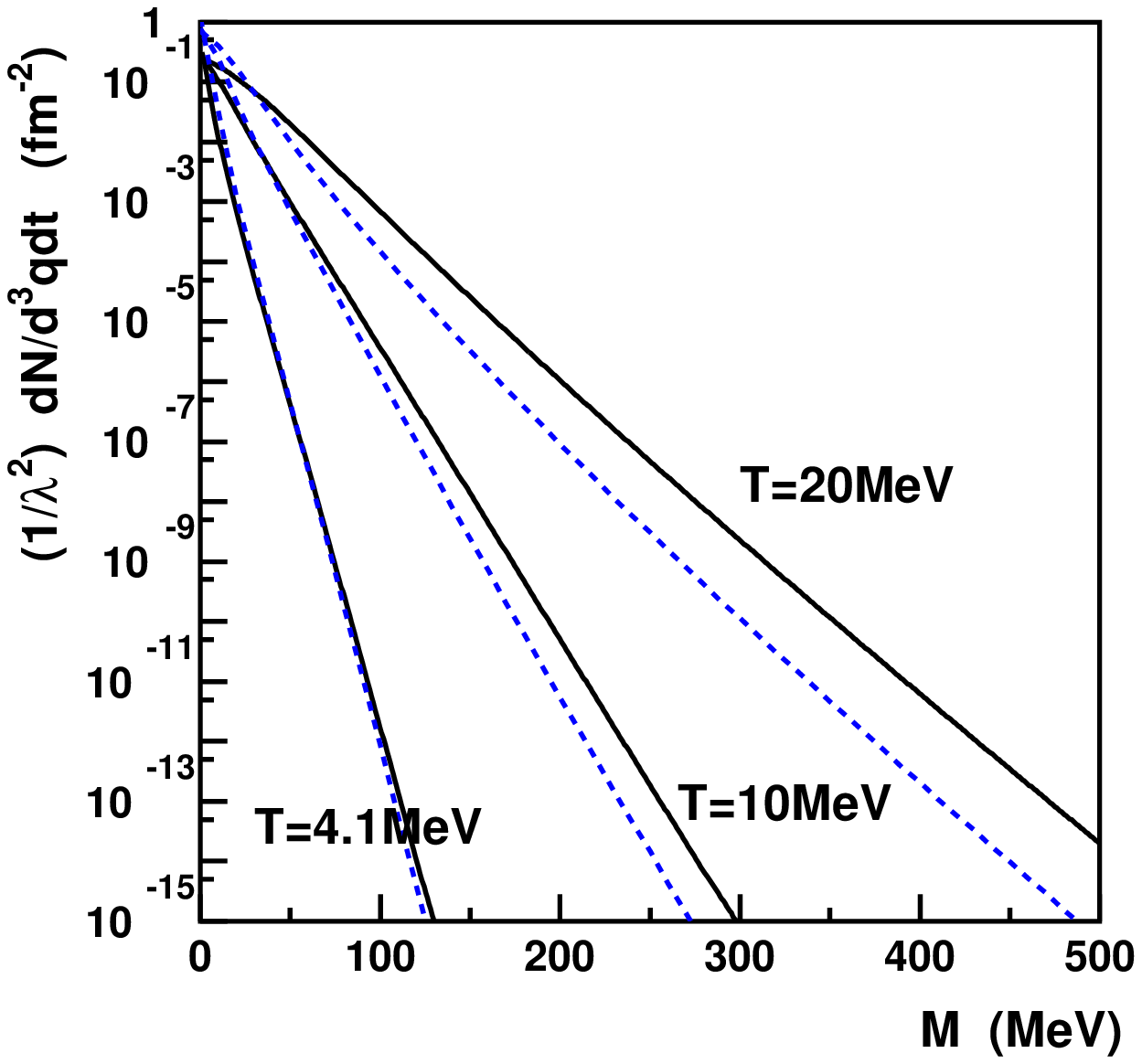,width=5cm,height=5cm,angle=0}}
\caption{ (left panel) Single particle width obtained from Eq. 
(\ref{ims}) in the quasi-particle T-matrix approximation (solid line) 
and from the Boltzmann collision integral (dashed line).
(right panel) Production rate of mesons as function 
of their mass 
in the quasi-particle T-matrix approximation.
 Quantum and  semiclassical production rates are represented by solid and 
dashed lines respectively.}\label{figqp}
\end{figure}
\noindent
where the disconnected two-particle propagator is~:
\begin{eqnarray}
\label{twpro}  &
<{\bf p}|{\cal G}^\pm({\bf P},\Omega)|{\bf p}^{'}> =  
(2 \pi)^3 \delta^3({\bf p}-{\bf p}^{'})\int \frac{d\omega^{'}}{2 \pi}
\int \frac{d\omega}{2 \pi} \big(
G^<({\bf P}/2+{\bf p},\omega-\omega^{'}) &
\nonumber \\ &
G^{<}({\bf P}/2-{\bf p},
\omega^{'})   -G^>({\bf P}/2+{\bf p},\omega-
\omega^{'})G^>({\bf P}/2-{\bf p},\omega^{'}) \big) 
/(\Omega -\omega \pm 
i\epsilon) \ . &
\end{eqnarray}
\begin{figure}
\parbox[t]{5.5cm}{
\psfig{figure=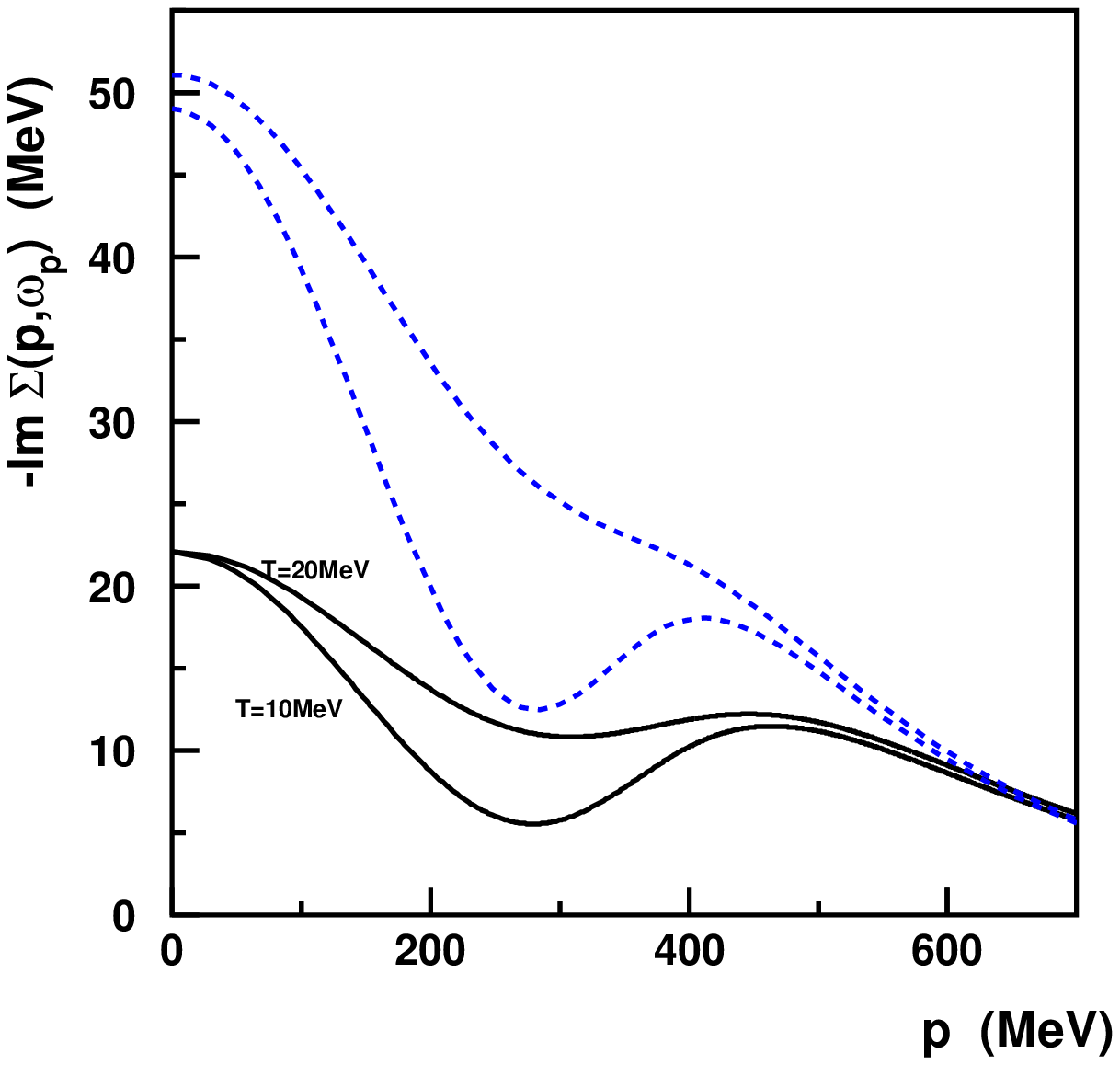,width=5cm,height=5cm,angle=0}}
\hspace{3ex}
\parbox[t]{5.5cm}{
\psfig{figure=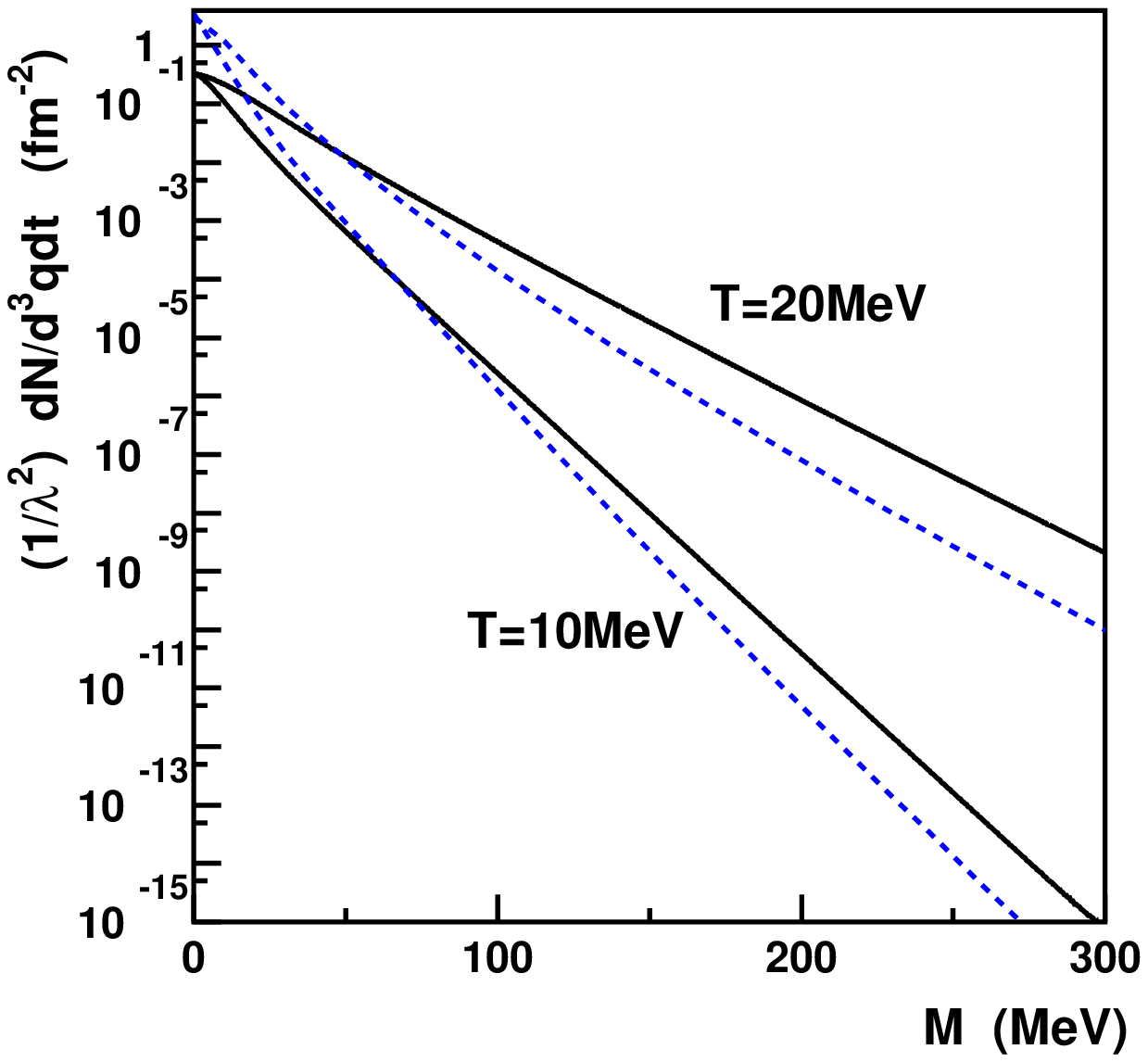,width=5cm,height=5cm,angle=0}}
\caption{\label{2} (left panel) Single particle width obtained from Eq. 
(\ref{ims}) in the self-consistent T-matrix approximation (solid line) 
and from the Boltzmann collision integral (dashed line).
(right panel) Production rate of mesons as function of their mass  
in the self-consistent T-matrix approximation.
 Quantum and  semiclassical production rates are represented by solid and 
dashed lines respectively.} \label{figself}
\end{figure}

The T-matrix equation has to be solved iteratively together with Eqs.
(\ref{iteq}).  The functional defining the self-energy is now
\begin{eqnarray}
\label{ims}
{\rm Im}\Sigma^+(p,\Omega)& = &\int\frac{d\omega}{2 \pi}\int 
\frac{d^3k}{(2 \pi)^3}
A(k,\omega)  \Big( f(\omega)+g(\omega+\Omega) \Big)
\nonumber \\ & &  <({\bf p}-{\bf k})/2|{\rm Im}T^+({\bf p}
+{\bf k},\Omega+\omega)|({\bf p}-{\bf k})/2>_A
\end{eqnarray}
instead of the Born approximation used in the previous section.
The quasiparticle approximation in the T-matrix ladder consists in
replacing the two fermion propagator (\ref{twpro}) by 
quasiparticle propagators \cite{rostock}.
First applications of the self-consistent T-matrix scheme to finite temperature
nuclear matter \cite{pb3} show important quantitative differences
between the quasiparticle and self-consistent T-matrix results. Both
at high temperatures where scattering is important and at lower temperatures 
close to the pairing transition.

The quasi-particle T-matrix approximation gives in-medium cross sections, but
all the nucleon propagators are on-shell in the ladder resummation. The
 single particle width and spectral function are the same as obtained 
using the Boltzmann collision integral with the cross section obtained 
from the in-medium T-matrix. Indeed the left panel in Fig. 3 shows the 
equivalence of the two calculations (the small differences are due to 
numerical inaccuracy and approximate treatment, angular averaging,
 of the energy conserving $\delta$ function in the collision term).
The semiclassical production rate is defined as in the Born approximation 
except that the two-particle potential is replaced by the in medium T-matrix.
The quantum production rate differs from the semiclassical one only by
the inclusion of 3-nucleon processes corresponding to having both propagators
 in the meson self-energy loop off-shell. This contribution is most 
important at large
meson energy and increases with temperature.

The self-consistent T-matrix, using off-shell nucleon propagators in the 
ladder, leads to different spectral function and single
particle energies \cite{pb3}. In a self-consistent calculation 
 there is no simple relation between 
the single-particle width obtained from the Boltzmann equation and from
the T-matrix scheme. As in the Born approximation we find that on-shell
propagation leads to larger relaxation rates. Part of this effect can be 
understood as due to the different densities of states for the two-particle
propagators in the quasi-particle and self-consistent T-matrix 
calculations \cite{dickhoff}. The optical theorem,
 if using the density of states corresponding 
to on-shell nucleons, leads to an overestimation of the cross-section.

The
 production rate is larger in the quantum calculation, especially
for energetic particles. The difference comes from the 3-body contribution 
discussed above and from the far off-shell regions of the spectral function.
As in the Born approximation, the spectral function obtained in the 
self-consistent iteration is larger (wider)  far from the shell. 
On the other hand, close to 
the quasi-particle pole the spectral function calculated from the Boltzmann 
collision term is wider, reflecting the systematically larger single particle
 width obtained with on-shell nucleons.

\section{Summary}
 We have calculated particle production in nonequilibrium quantum transport
framework. The production rate could not be defined in the initial 
stage of the evolution. It is possible only at large times, when the system 
is already equilibrated. The corresponding rates can be calculated using 
equilibrium spectral functions, obtained in an iterative solution of 
coupled equations (\ref{iteq}).
 The quantum production rate is larger than the semiclassical
one for energetic particles. This can be understood as due to 
3 and sequential 
$n$-nucleon collisions. In-medium T-matrix calculations allow to include 
in-medium modifications of the cross section in the picture. The
cross section for Boltzmann type collision integrals and for the semiclassical
production rates has to be modified by a factor describing a different density 
of states \cite{pn}. After such a renormalization of the cross sections the 
relaxation rates would be similar in the quantum and semiclassical transport
equations, but the difference between the quantum and semiclassical 
production rates would increase.

\section*{Acknowledgments}
This work was partly supported by the National Science Foundation
under Grant PHY-9605207.


\section*{References}
\begin{thebibliography}{99}
\bibitem{pawel1} P. Danielewicz, {\em Ann. Phys. (N.Y.)} {\bf 152} (1984) 305.
\bibitem{pn} P. Nozi\`ere, {\it Theory of Interacting Fermi Systems},
(New York, Benjamin, 1994)
\bibitem{pb1} P. Bo\.zek, {\em Phys. Rev.} {\bf C56} (1997) 1452.
\bibitem{pb2} P. Bo\.zek, in Proceedings of the I-V workshop on {\it
Nonequilibrium physics at short time scales}, ed. by K. Morawetz, P. Lipavsk\'y
and V. \v{S}pi\v{c}ka, (Universit\"at Rostock, Rostock, 1998), nucl-th/9807068.
\bibitem{haus} R. Hausmann, {\em Phys. Rev.} {\bf B49} (1994) 12975.
\bibitem{rostock} T. Alm, G. R\"opke, A. Schnell N.H. Kwong and S. K\"ohler,
{\em Phys. Rev.} {\bf C53} (1996) 2181.
\bibitem{pb3} P. Bo\.zek, {\em Phys. Rev.} {\bf C59} (1999) 2619.
\bibitem{dickhoff} W.H. Dickhoff, {\em Phys. Rev.} {\bf C58} (1998) 2807.
\end{thebibliography}
\end{document}